# Microresonator Brillouin Laser Stabilization Using a Microfabricated Rubidium Cell

William Loh[1,2*], Matthew T. Hummon[1*], Holly F. Leopardi[1], Tara M. Fortier[1], Frank Quinlan[1], John Kitching[1], Scott B. Papp[1], Scott A. Diddams[1]

[1]*National Institute of Standards and Technology, 325 Broadway, Boulder, Colorado 80305, USA*
[2]*Current address: MIT Lincoln Laboratory, 244 Wood Street, Lexington, Massachusetts 02420, USA*
*These authors contributed equally to this work
Email: Scott.Diddams@nist.gov

**Abstract:** We frequency stabilize the output of a miniature stimulated Brillouin scattering (SBS) laser to rubidium atoms in a microfabricated cell to realize a laser system with frequency stability at the $10^{-11}$ level over seven decades in averaging time. In addition, our system has the advantages of robustness, low cost and the potential for integration that would lead to still further miniaturization. The SBS laser operating at 1560 nm exhibits a spectral linewidth of 820 Hz, but its frequency drifts over a few MHz on the 1 hour timescale. By locking the second harmonic of the SBS laser to the Rb reference, we reduce this drift by a factor of $10^3$ to the level of a few kHz over the course of an hour. For our combined SBS and Rb laser system, we measure a frequency noise of $4\times10^4$ Hz$^2$/Hz at 10 Hz offset frequency which rapidly rolls off to a level of 0.2 Hz$^2$/Hz at 100 kHz offset. The corresponding Allan deviation is $\leq 2\times10^{-11}$ for averaging times spanning $10^{-4}$ to $10^3$ s. By optically dividing the signal of the laser down to microwave frequencies, we generate an RF signal at 2 GHz with phase noise at the level of -76 dBc/Hz and -140 dBc/Hz at offset frequencies of 10 Hz and 10 kHz, respectively.

## 1. Introduction

Whispering-gallery mode microresonators [1−6] provide a key technological innovation that allows for low optical loss in a tightly-confined resonator geometry, all within a device footprint of a few millimeters or less. With the continual reduction of resonator loss over recent years, microresonators have now exceeded quality factors (Q) of $10^{10}$ [3]. At these extremely high levels of Q, the intracavity optical power is enhanced by several orders of magnitude and reaches a point where nonlinear optics can be efficiently excited using only microwatts of input power [7−9]. Recently, this technology has been utilized to create Kerr nonlinear frequency combs [7, 10−13], as well as narrow-linewidth lasers based on injection locking [14], stimulated Brillouin scattering (SBS) [9, 15−17], and stimulated Raman scattering (SRS) [8, 18]. Moreover, these devices have been demonstrated in a variety of platforms ranging from the ultrahigh-Q resonators (>$10^{10}$) in $CaF_2$ [3] to the chip-integrated resonators in $SiO_2$ [1, 16] and $Si_3N_4$ [19]. The development of microresonator SBS lasers in particular have resulted in fundamental frequency noise floors <0.1 $Hz^2$/Hz [9, 16, 20] and operating laser linewidths below 240 Hz [20] within a standalone device or below 90 Hz [17] when stabilized to a second passive microresonator. When compared to conventional semiconductor lasers exhibiting linewidths in the range of 10 kHz – 1 MHz, the SBS lasers are able to achieve orders of magnitude superior performance while maintaining a similar size scale.

Despite these remarkable characteristics, one common challenge that all microresonator devices currently face is that the resonant light propagates in dielectric materials for which the combined thermal-expansion and thermo-optic coefficients are typically nonzero. Thus, ambient temperature fluctuations, as well as internal optical power variations couple directly to the microresonator temperature and its resonant frequencies [21−23]. As an example, for silica glass, an environmentally-induced change of the resonator temperature by 1 mK would lead to an optical frequency shift of ~1 MHz. Similarly, the large circulating powers within the microresonator cavity combine with weak residual absorption to give temperature shifts that ultimately result in noise and drift of the cavity resonances of comparable order of magnitude. This temperature-driven drift impedes the ability of microresonator devices (i.e., the parametric frequency comb, Brillouin laser, etc.) to hold a constant frequency over time. Previous work on stabilizing this drift utilized the difference in temperature sensitivity between two microresonator modes to measure the variation in temperature as a relative displacement between the two modes [24−26]. In addition, narrow transitions in atomic or molecular systems have historically been used to provide long-term stability for laser oscillators, but typical implementations, even well engineered compact ones, employ bulk optical components and centimeter-scale vapor cells [27−29]. More recently, microresonator based lasers and parametric oscillators have been locked to rubidium atoms at 780 nm, providing optical stability as good as ~2 kHz [30−32].

In this work, we move beyond previous approaches by employing a microfabricated rubidium cell having external dimensions <0.1 $cm^3$ to provide long-term frequency stability to the low-noise SBS laser. Such microfabricated cells are a critical component of chip-scale atomic clocks operating at microwave frequencies [33], but their properties for Doppler-free saturated absorption spectroscopy in the optical domain have only been explored in preliminary studies [34]. Vertical cavity surface-emitting lasers have been locked to Doppler-broadened lines in microfabricated vapor cells and long-term instabilities in the optical frequency in the range of $10^{-9}$ have been achieved [35]. Here we demonstrate the frequency locking of a 6-mm diameter microresonator SBS laser operating at 1560 nm (linewidth = 820 Hz) to the D2 transition of the Rb cell at 780 nm, which improves the SBS laser drift from the level of a few megahertz to ~5 kHz at the 1000 s time scale. Significantly, the architecture we employ for the generation, spectroscopy and stabilization of the SBS laser includes components that are all amenable to further chip-scale integration [36, 37]. Finally, through frequency division [38−40] of our optical signal down to 2 GHz, we further show the ability to generate RF signals with a phase noise level and frequency stability superior to many high-performance RF oscillators.

2. **System Configuration**

The diagram of our combined system including both the SBS laser and the Rb cell is illustrated in Fig. 1. The overall system implementation first stabilizes a pump laser at 1560 nm to a microrod resonator for the generation of a low-noise SBS wave in the counter-propagating direction [41]. The SBS light is then frequency doubled to reach 780 nm, and the resulting signal is locked to the D2 transition of Rb atoms in a miniature cell, which establishes the long-term stability of the laser.

The SBS laser subsystem comprises a semiconductor planar lightwave circuit pump laser operating at 1560 nm wavelength whose output is first sent through a $LiNbO_3$ phase modulator (PM) and next into a semiconductor optical amplifier (SOA) to increase the total optical power to ~30 mW. The output of the SOA is then sent into a circulator, which redirects the light into a 6-mm diameter microrod resonator [6] to generate SBS lasing. The transmitted power from the microresonator including the phase modulation sidebands is photodetected and subsequently mixed with the local oscillator (LO) previously used as the driver for phase modulation. This generates the error signal that we utilize to Pound-Drever-Hall lock the pump laser to the microresonator resonance via feedback on the laser current. The counter-propagating SBS signal generated within the microresonator enters the circulator in the reverse direction and becomes redirected to our rubidium stabilization system.

In the rubidium subsystem, the SBS signal is first amplified by a SOA to ~30 mW and then subsequently frequency doubled to 780 nm in a waveguide periodically-poled lithium niobate crystal (PPLN). The free space output of the doubled light is routed through a polarizer (Pol) and into a 780 nm $LiNbO_3$ phase modulator, which generates the sidebands for stabilization of the probe beam onto the rubidium D2 transition. We note that a fiberized $LiNbO_3$ phase modulator was used here, and thus the free-space light was interfaced to the fiber connectors at both the input and output of the modulator. At the modulator output, the light is directed into a 90:10 splitter where 10% of the light is monitored on a Si photodetector for stabilization of the intensity and residual amplitude modulation (RAM) of the probe beam. The rest (90%) of the light is sent to the miniature Rb cell, which comprises an inline polarizer followed by the Rb sample. The light transmitted through the Rb cell is then sent to a second Si photodetector with an attached reflector that reflects 70% of the optical power. This reflected light provides the pump beam for the Doppler-free saturated absorption signal in Rb, while the transmitted 30% is mixed down to generate an error signal to lock the SBS laser to the cell. The feedback is applied to the SOA in the SBS laser system which changes the optical power and thus the heating of the microrod resonator. Since the pump laser is locked to the microrod resonance, the heating or cooling of the microresonator provides direct tuning of the SBS signal for stabilization onto the Rb transition.

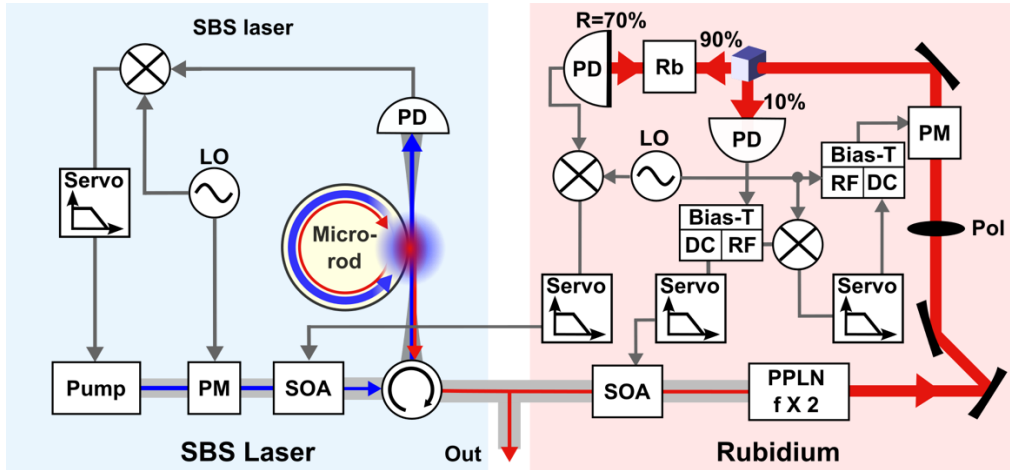

Fig. 1. System diagram of the combined laser consisting of a SBS laser locked to a miniature rubidium cell.

## A. SBS Laser Characterization

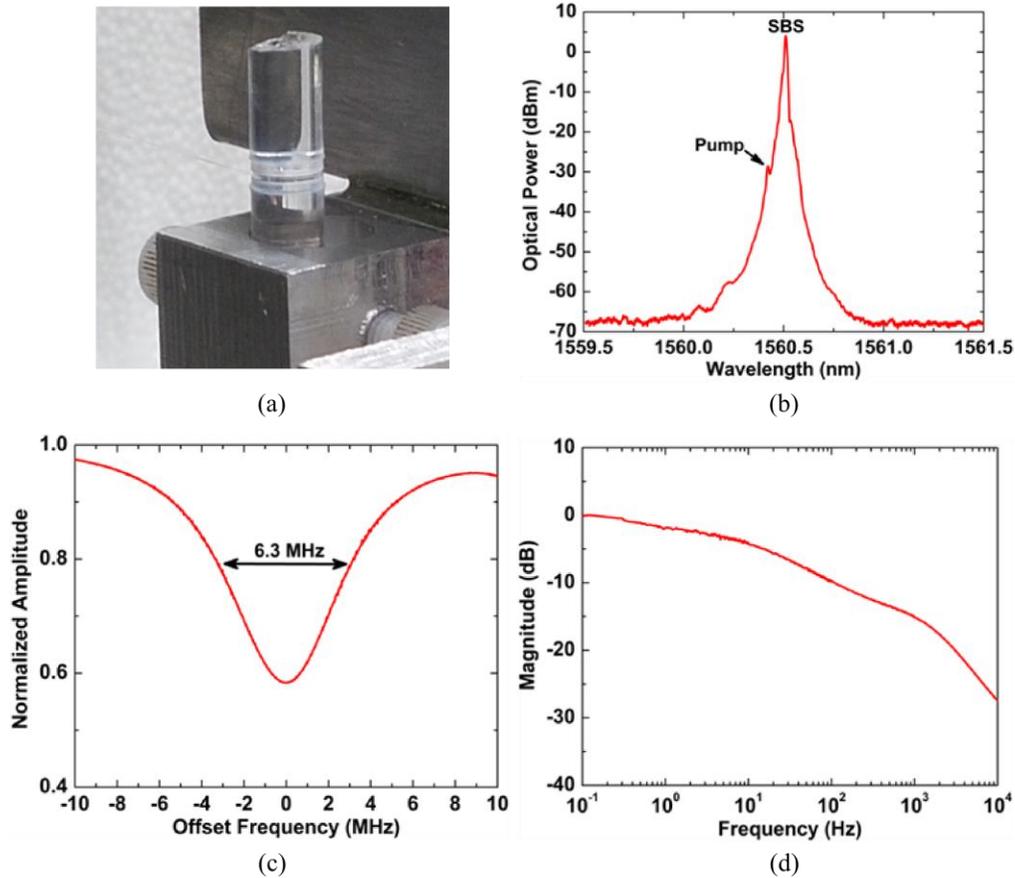

Fig. 2. Characterization of the SBS laser. (a) Photograph of the SBS microrod resonator coupled to a tapered fiber. The microrod diameter is 6 mm. (b) Spectrometer resolution-limited measurement of the SBS laser optical spectrum showing 2.5 mW SBS output power. (c) Plot of the microrod mode at 1560 nm under low optical pump powers indicating a mode linewidth of 6.3 MHz. (d) Frequency response characterizing the ability to tune the SBS laser frequency via modulation of the optical power. The bandwidth of the response is 4.5 Hz.

A photograph of our microrod resonator used for the generation of the SBS signal is shown in Fig. 2a. The microrod diameter is 6 mm and is positioned next to a tapered optical fiber for the coupling of light into and out of the microresonator. A resistive heater, located under the aluminum microrod mount, provides coarse tuning of the microrod resonance frequency. Figure 2b shows the optical spectrum centered at 1560 nm generated from the SBS laser in the counter-propagating direction. 2.5 mW of SBS power is obtained from a total of ~10 mW power input into the microrod. However, a residual signal of the pump also remains due to unwanted backscatter within the microresonator at a level of -32.5 dB relative to the SBS signal. This backscattered pump signal is separated from the SBS signal by 11 GHz and can be filtered out with the use of conventional filters if necessary.

Figure 2c shows the mode profile of the SBS microresonator as measured in the transmitted port. The measurement was taken at low optical pump powers as to avoid thermal broadening of the mode. Furthermore, we calibrated the frequency axis by applying modulation sidebands at a known frequency and comparing with the resulting separation of the sidebands on the oscilloscope trace. As a result of the specific wavelength required to access the 780 nm Rb transition, the selection of suitable microresonator modes becomes limited to those that when pumped yield a SBS signal that reaches the Rb D2 wavelength. A resistive heater is used to provide tuning of the microrod resonance frequencies, which provides greater freedom in the choice of modes. The microrod mode we used exhibits a linewidth of 6.3 MHz. In general, this achieved linewidth is broad for a microrod mode as we have demonstrated linewidths on the order of 200 kHz − 300 kHz in previous microrod cavities [6, 17]. Note that a nearby mode on the higher frequency side causes the mode profile to appear slightly distorted.

Figure 2d depicts the frequency response characterizing the ability of an applied amplitude modulation on the pump signal to cause a change in the frequency of the SBS laser. As a result, the frequency response of Fig. 2d is a measure of the speed that we can tune the SBS laser and thus provides information on the bandwidth of our lock to the Rb cell. We measure this 3-dB bandwidth to be ~4.5 Hz. Note that the slow response of our frequency tuning presents no issue here as our ideal Rb lock should preserve the SBS laser's noise at higher offset frequencies (> 10 Hz) since the SBS frequency noise there is much lower than the equivalent noise achieved in the Rb spectroscopy. Only at low offset frequencies should the Rb lock take effect and stabilize the drift of the SBS laser.

B. *Stabilization to Miniature Rb Cell*

A photograph of the miniature rubidium vapor cell is shown in Figure 3a. The vapor cell consists of a MEMS fabricated 2-mm thick silicon frame to which pyrex windows are anodically bonded. The frame contains two chambers connected by a narrow channel, a small chamber for an alkali metal dispenser pill and a larger chamber with a 3 mm × 3 mm clear aperture for the probe laser. The first pyrex window is anodically bonded to the frame in air, after which the dispenser pill is inserted into the small chamber. The cell is then inserted into a vacuum chamber where the second window is anodically bonded to the cell at a temperature of 300 °C in an atmosphere with residual pressure of ~$10^{-7}$ torr. Following completion of the bonding, activation of the rubidium dispenser pill is achieved via optical heating of the pill using a 2 W laser at 975 nm with exposure time of about 10 s [42]. The vapor cell contains both isotopes $^{85}$Rb and $^{87}$Rb in their natural abundance. The vapor cell is then diced to its final dimensions of 5 mm x 7 mm. A chip thermistor and a set of four resistive chip heaters are epoxied to the sides of the vapor cell to provide active temperature stabilization of the vapor cell. A thin linear film polarizer is epoxied to the input aperture of the cell to provide a fixed probe laser polarization.

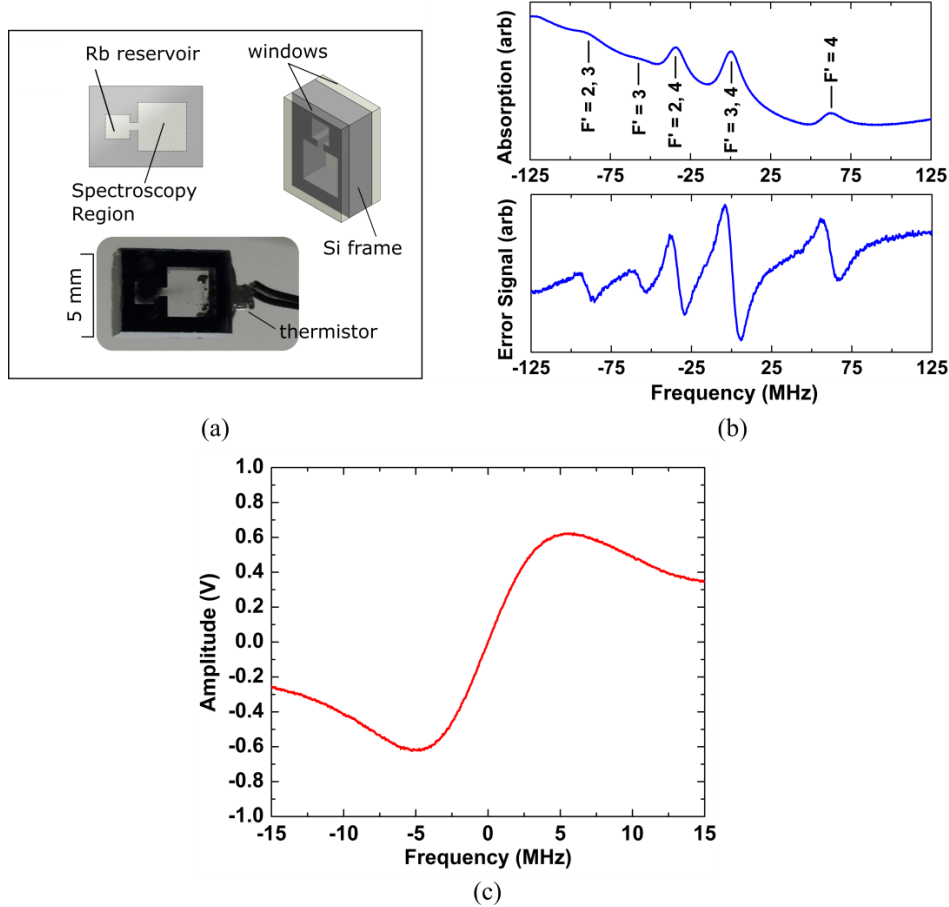

Fig. 3. Characterization of the Rb reference cell. (a) Schematic and photograph of the 5 mm × 7 mm Rb cell. The windows are anodically bonded to the silicon frame. The photograph shows the cell with a thermistor epoxied to its base. (b) 780 nm saturated absorption spectrum and error signal of $^{85}$Rb taken by scanning the pump laser across the F = 3 manifold. (c) Spectroscopic error signal with 32-point averaging generated by scanning the SBS laser over the F=3 to F'=3,4 $^{85}$Rb D2 transition. The SBS laser was locked to this transition for all work described in this paper.

Stabilization of the SBS laser to the Rb D2 transition is achieved using Doppler-free FM spectroscopy. The probe beam is phase modulated at a modulation depth of ~1 to produce frequency sidebands at +/- 7 MHz. When the probe beam enters the Rb cell it has a diameter of 3 mm and contains ~300 µW of optical power at 780 nm. To achieve an optimal signal-to-noise ratio for the error signal, we operate the Rb cell at a temperature of 68 °C. The Rb cell is tilted away from normal laser beam incident to avoid unwanted etalons. Figure 3b shows the saturated absorption spectrum and corresponding error signal taken using the miniature vapor cell. The spectra were taken by scanning the pump laser across the resonances, where we determine the sub-Doppler resonances to have linewidths of ~10 MHz. The measured linewidths are slightly larger than the natural linewidth of the transition of 6 MHz, and the residual broadening is likely due to power broadening. Figure 3c shows the error signal generated as the SBS laser is tuned across the F=3 to F'=3, 4 cross-over peak in $^{85}$Rb. As we will see in Fig. 4, the Allan deviation generated from the Rb lock does not completely average down over long time scales, and thus the frequency noise is not entirely white. However, as we show next, we believe this is a result of technical noise in our system and not due to a fundamental limit of the Rb atoms. If we assume that we are able to reach the white-noise limit of our Rb lock, we use the error signal slope of 0.2 V/MHz to calculate a frequency noise floor of 174 Hz/√Hz, which yields an Allan deviation of $4.5 \times 10^{-13}$ at 1 s limited by a combination of electronic noise and shot noise. Our projected value of Allan deviation using a microfabricated Rb cell compares favorably to previously demonstrated Rb spectroscopy experiments in bulk Rb cells [43], which demonstrated stability in the range of $10^{-12}$ at 1 s.

To monitor the long term stability of the SBS laser stabilized to the Rb atoms in the miniature cell we measure the optical heterodyne frequency between the SBS laser and an independent cavity-stabilized optical frequency comb, as described in the following section. To achieve long term stability of the SBS laser lock point, active control of several parameters of the Rb cell and probe laser system are required. First, temperature stabilization of the Rb cell is required to maintain a constant rubidium vapor pressure in the cell. Fluctuations of the cell temperature lead to varying Rb density, causing variations in average absorption of the probe beam in the Rb cell. This probe beam intensity fluctuation then causes shifts in the lock point due to the AC Stark effect. Additionally, we actively stabilize the intensity and RAM [44] of the probe beam directly before the Rb cell via a beam pick-off and a single photodiode (see Fig. 1). Variation of the RAM of the probe beam leads to voltage offsets of the error signal, thus shifting the lockpoint away from the line center. A bias-tee splits the photodetected signal into DC and RF components. The DC component is used to stabilize the probe beam intensity via feedback to the SOA drive current. The RF component is used to monitor and stabilize the probe beam RAM. The RAM at 7 MHz arises due to differential phase shifts caused by the natural birefringence of the LiNbO$_3$ crystal in the phase modulator, which results in amplitude modulation when the input light's polarization is mismatched to the polarization axes of the crystal. A DC bias voltage can be applied to the phase modulator to cancel these phase shifts and suppress the RAM [45]. We monitor the in-phase component of the RAM via demodulation with the local oscillator at 7 MHz and use the resulting signal to servo the RAM via the DC bias voltage to the phase modulator. Since the birefringence of the phase modulator is also temperature dependent, we also actively stabilize the temperature of the phase modulator using a thermo-electric cooler. In the absence of any active RAM stabilization, we observe RAM fluctuation on the order $10^{-2}$, leading to shifts of the SBS laser on the order of 100's of kHz. Using this stabilization technique, we achieve fractional RAM stability of about $10^{-4}$. At this level, frequency shifts due to RAM fluctuations are no longer the dominant source of instability in our SBS-Rb laser system.

Table 1 summarizes our analysis of the factors that currently limit our laser stability. We report this analysis for frequency shifts at 1560nm, and we note that the corresponding shift at 780 nm is a factor of two larger. The sensitivity of the Rb lock frequency to fluctuations of the laser power is 5 kHz/μW, which given the stability of optical power at the level of 0.5 μW in our system, accounts for an expected frequency instability of ~2 kHz. Similarly, from temperature sensor measurements, we determine that cell temperature fluctuations are stabilized to within 0.3 °C, yielding a frequency instability of ~10 kHz for the Rb lock. We measure magnetic fields to be passively stable at the milligauss level near the Rb cell. Therefore, from our analysis, we find that variations of the cell temperature are currently the dominant noise process that limits the frequency stability of the combined SBS / Rb cell system. We believe that it should be possible to further reduce these systematic effects with improved engineering in order to reach the ideal white-noise limit calculated earlier.

Table 1. Frequency stability analysis of the combined SBS and Rb laser system.

| Parameter | Frequency shift sensitivity | Stability | Expected instability |
|---|---|---|---|
| **Laser power** | 5 kHz/μW | 0.5 μW | 2 kHz |
| **Cell temperature** | 30 kHz/ °C | 0.3 °C | 10 kHz |
| **Magnetic field** | 1 kHz/mG | 1 mG | 1 kHz |
| **RAM** | 20 MHz/fractional RAM | $10^{-4}$ | 2 kHz |

### 3. Measurement Results

In this section, we describe the result of stabilizing the SBS laser to the Rb cell, demonstrating a laser that exhibits excellent frequency-noise performance at both short- and long-term time scales. Even though we stabilize to the cell at 780 nm, all measurements are performed using the output before the frequency doubler at 1560nm. Figure 4a shows the frequency noise of the three laser systems used in our study, which includes the pump laser, the SBS laser, and the SBS laser locked to the Rb cell. At low offset frequencies (<1 kHz), the measurements were taken by direct heterodyne against a 1550nm cavity-stabilized laser that has ~1Hz linewidth and frequency stability <3x10$^{-13}$ for averaging times up to a 1000 s. To reach our Rb-stabilized SBS laser at 1560 nm, we first electro-optically modulate and then nonlinearly broaden the spectrum of the cavity-stabilized laser to achieve a 10 GHz repetition rate comb with a span of >20 nm [46]. At higher offset frequencies (>1 kHz), the SBS laser noise surpasses that of the cavity-stabilized laser system, and we instead use a Mach-Zehnder interferometer of 200 m delay length to measure the frequency noise. These two measurement techniques are combined to yield the overall noise spectrum of Fig. 4a. Our results indicate the pump frequency noise to be an order of magnitude or more above that of the SBS laser; however, at lower offset frequencies, the SBS noise also increases due to its sensitivity to temperature fluctuations [47, 48]. For frequencies below 10 Hz, we stabilize the SBS laser to the Rb atoms, which acts as a reference to compare to and correct for the frequency excursions of the SBS laser. From Fig. 4a, we find the improvement in frequency noise to be >10 dB at 1 Hz offset frequency. We further integrate the frequency noise spectral density of our combined laser system [49] and estimate our laser linewidth to be 820 Hz. This SBS laser linewidth is a factor of 3−4 times larger than the linewidth of our SBS laser reported in Ref. [20] and is due to the limitations of having access to only lower-Q modes near the Rb transition.

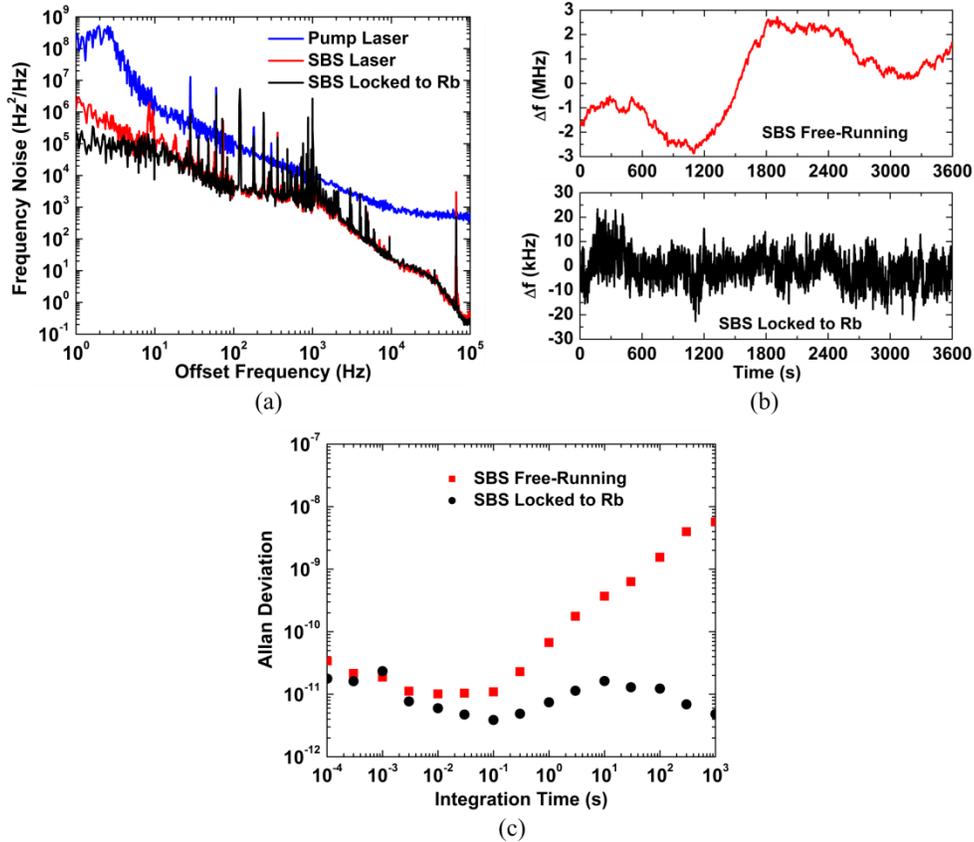

Fig. 4. Measurements of the combined SBS and Rb laser system. (a) Frequency noise of the pump laser (blue), SBS laser (red), and SBS laser locked to Rb (black). (b) 1-hour time record of the free-running (upper) and Rb-locked (lower) SBS laser. Both sets of data are offset from the nominal optical frequency of ~192 THz. (c) Allan deviation of the free-running and locked SBS laser.

The measurement of the SBS laser's frequency excursions over time provides a separate assessment of the laser's frequency stability. Figure 4b shows a time record of the SBS laser's frequency, over the course of one hour, as measured from a heterodyne against the cavity stabilized laser and frequency comb reference. When free-running, the total frequency shift of the SBS laser is 5.5 MHz for the duration of the measurement. However, when locked to the miniature Rb cell, we find the laser's frequency to be bounded within a range of ~30 kHz. The rapid fluctuations in frequency are due to laser noise at higher offset frequencies, while the slow movement of the center frequency characterizes the laser's drift over time. As can be observed from Fig. 4b, the long-term frequency excursions of the system are confined within the range of a few kilohertz over the one-hour measurement.

We further analyze the SBS laser's frequency stability by computing an Allan deviation of the frequency fluctuations as a function of the integration time. We are specifically interested at both short and long time scales and thus compute the Allan deviation for measurements ranging from 1 – 1000 seconds in duration and for gate times ranging from

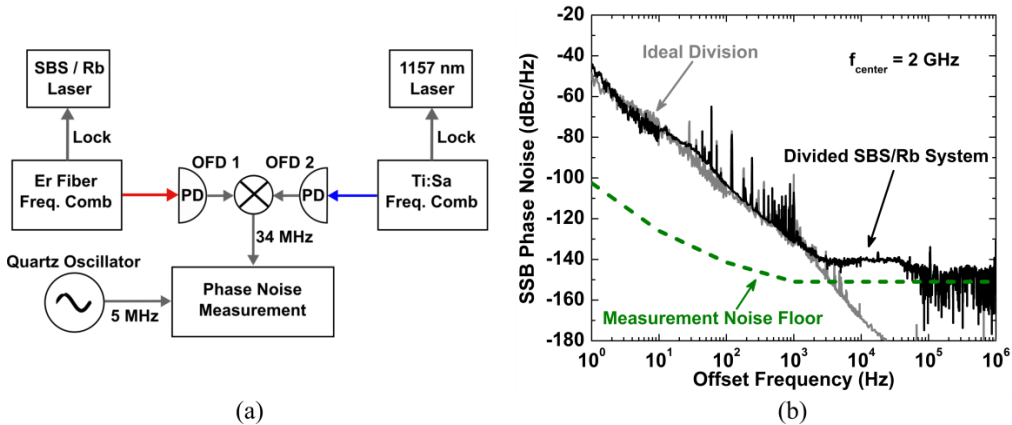

Fig. 5. Optical frequency division of SBS and Rb laser signal down to 2 GHz. (a) Diagram of the system used for phase-noise measurement of the divided SBS signal. (b) Phase-noise spectrum of the SBS/Rb laser system divided down to 2 GHz. The phase noise of the quartz-referenced measurement system (green dashed line) is provided. The expected phase noise resulting from ideal division of the frequency noise (gray line) is also indicated.

100 μs to 100 ms. The various results are combined to form the overall Allan deviation plot of Fig. 4c. We find that at short integration times below 0.01 s, the Allan deviation of the free running and locked SBS lasers converge to nearly the same value. This is expected since for offset frequencies above ~10 Hz in Fig. 4a, the noise of the free-running and locked SBS lasers are identical. However, at longer time scales, the Rb cell reference maintains the Allan deviation at a level near $10^{-11}$, averaging down slightly over time for time scales above 10 s. For the case of the free-running SBS laser, the frequency drift of the laser causes the Allan deviation to increase sharply at long time scales, reaching the level of $10^{-8}$ at 1000 seconds of integration. At these longer time scales, we find the stability improvement resulting from the Rb lock to be ~3 orders of magnitude.

The excellent performance demonstrated at 1560 nm by the combined SBS laser and Rb cell system can be coherently transferred to a RF frequency through the process of optical frequency division (OFD) [38−40]. Here, we generate a 2 GHz RF signal by dividing down the stabilized 192 THz light from our locked SBS laser using an Er-doped fiber laser optical frequency comb (Fig. 5a). This optical frequency division was accomplished by locking the nearest comb line of a self-referenced Er optical frequency comb to the SBS/Rb laser system via feedback on the comb's repetition rate. In this manner, the frequency fluctuations of the SBS laser are transferred to fluctuations on the comb repetition rate, but divided down by the ratio of the optical frequency to the microwave frequency. For a comb repetition rate of ~155 MHz, we photodetect the comb and choose the 13$^{th}$ harmonic at 2 GHz filtering out all other components. We compare the generated 2 GHz RF signal to the 2 GHz signal produced by a second self-referenced Ti:S optical frequency divider locked to a cavity-stabilized laser at 1157 nm (Allan deviation $10^{-16}$ at 1s). The two RF signals are mixed and down-converted to a new microwave signal at 34 MHz that contains the information for the frequency fluctuations of the SBS laser. We then measure the noise on this signal using a quartz-referenced phase noise measurement system to generate the plot of Fig. 5b. For a 2 GHz carrier, we achieve a phase noise of -76 dBc/Hz at 10 Hz offset frequency, which rolls off to a value of -141 dBc/Hz near 3 kHz offset. At higher offset frequencies beyond 100 kHz, the measurement reaches the noise floor limit of the quartz-referenced measurement system. Additionally, we note that recent demonstrations of self-referenced microcombs [50, 51] suggest that in the future an entire optical frequency division system, including the SBS laser and Rb reference discussed here, may be possible in a compact package.

## 4. Summary

We have shown that the SBS laser, which offers excellent short-term noise performance but exhibits frequency drift at long time scales, can be synergistically combined with a vapor cell of Rb atoms to form a system that achieves low-noise performance at any time scale of interest. By miniaturizing the Rb cell to dimensions of <0.1 cm$^3$, the combined laser system demonstrates the potential for future integration into a compact centimeter-scale package, requiring 100−500 mW power, but with frequency noise and stability surpassing that of both fiber and solid state lasers. We further demonstrated that via optical frequency division, the SBS / Rb laser can be used to generate stable microwave signals with lower noise than that of many high-performance RF oscillators and reaches -76 dBc/Hz at 10 Hz offset frequency. Currently, the use of the Rb reference cell limits the applicability of this frequency stabilization scheme to the Rb transition wavelength of 780 nm or to multiples of 780 nm utilizing frequency multiplication. However, with the eventual development of octave-spanning chip-integrated microresonator frequency combs, we expect to be able to access any atomic or molecular frequency reference in the visible to near-infrared.

**Acknowledgments**

We thank Wei Zhang and Roger Brown for their valuable comments on this manuscript. We thank Susan Schima for fabrication help. This work was funded by NIST and the DARPA PULSE Program. WL acknowledges support from the NRC/NAS. This work is a contribution of the US Government and is not subject to copyright in the US. Mention of specific trade names is for technical information only and does not constitute an endorsement by NIST